\documentclass[%
 reprint,
superscriptaddress,
 amsmath,amssymb,
 aps,
]{revtex4-1}
\usepackage{graphicx}
\usepackage{dcolumn}
\usepackage{bm}
\usepackage{subfigure}
\usepackage{textcomp}

\newcommand\refeq[1]{Eq. \eqref{#1}}
\newcommand\reffig[1]{Fig. \ref{#1}}

\begin{document}

\preprint{APS/123-QED}

\title{A rational use of BCA code MARLOWE for ballistic effects of ion beam irradiation in the ion mixing formalism: comparison to Molecular Dynamics}%

\author{G. Demange}
 \email{gilles.demange@univ-rouen.fr}
 \affiliation{DEN/MDN/SRMA/LA2M, CEA Saclay, F-91191 Gif-sur-Yvette, France}
\author{E. Antoshchenkova}%
 \email{ekaterina.antoshchenkova@polytechnique.edu}
\affiliation{DEN/MDN/SRMA/LA2M, CEA Saclay, F-91191 Gif-sur-Yvette, France}
\author{M. Hayoun}%
\affiliation{LSI, \'Ecole Polytechnique, CNRS, CEA Saclay, Universit\'e Paris-Saclay, F-91128 Palaiseau, France}
\author{L. Lun\'eville}
\affiliation{DEN/SERMA/LLPR, CEA Saclay, F-91191 Gif sur Yvette, France}%
\author{D. Simeone}%
\affiliation{DEN/MDN/SRMA/LA2M, CEA Saclay, F-91191 Gif-sur-Yvette, France}
\date{\today}

\begin{abstract}

Understanding ballistic effects caused by ion beam irradiation, and linking them with induced structure can be a key point for controlling and predicting the microstructure of irradiated materials. For this purpose, we have investigated ballistic effects from an ion mixing formalism point of view. The displacement cascades in copper and AgCu alloy were obtained using binary collision approximation (BCA) and molecular dynamics (MD) simulations. We employed BCA-based code MARLOWE for its ability to simulate high energy displacement cascades. A first set of simulations was performed using both methods on pure copper for energies ranging from 0.5 keV to 20 keV. The results of BCA and MD simulations are compared, evidencing rationally parametrized MARLOWE to be predictive. A second set of simulations was then carried out using BCA only. Following experimental studies, AgCu alloy was subjected to 1 MeV krypton ions. MARLOW simulations are found to be in good agreement with experimental results.

\end{abstract}

\pacs{Valid PACS appear here}
\maketitle

\section{Introduction}
\label{sec:intro}

Patterned microstructures have recently become of particular scientific interest \cite{simeone2006}, as they could provide industry with materials of desired physical properties \cite{cheng1990}. For that purpose, ion beam mixing seems particularly promising to fabricate materials of desired patterned microstructure \cite{cross1993,bernas2003}. However, predicting and controlling the microstructure of irradiated materials still remains an extremely challenging issue. This firstly relates to the difficulty to estimate ballistic effects of displacement cascades, identified as the main driving force in patterning occurrence \cite{bellontheo}. The general purpose of this study is to compute these ballistic effects, providing with a first step toward microstructure control. 

This requires to simulate displacement cascades initiated by ion irradiation. This is usually achieved using molecular dynamics (MD) \cite{wiley}. Yet, this approach suffers from limitations, as it can barely handle more than $10^7$ atoms dynamics on massive parallel computing facilities, restraining incident ion energy to a few hundred keV \cite{Nordlund1995,Stoller2000,Terentyev2006,zarkadoula2013nature}. Still, previous studies suggested that incident energies up to a few MeV were required to achieve stationary patterning in binary alloys \cite{Krasnochtchekov2005,Chee2010}, so that this kind of problems is currently beyond reaching for MD. 

It was suggested in \cite{martin1984}, that this technological obstacle in the computation of ballistic effects could be skirted, using the ion mixing formalism \cite{sigmund1981}. This framework relies upon the ~idea that displacement cascades act on a very different time scale from the microstructure evolution. The microstructure is hence mainly sensitive to averaged effects of multiple cascades covering one another. As a result, the spatial heterogeneity of displacement cascades due to highly perturbed zones called subcascades, can be passed over. From this point, ballistic effects can be characterized by the sole mean of average features of displacement cascades: the atomic relocation frequency $\Gamma$, and the probability distribution $p_R$ for impinged atom to be relocated at a distance $r$ \cite{bellontheo}. In particular, particles individual trajectories within the cascade can be passed over. From this observation arose the idea to turn towards the  binary collision approximation framework \cite{robinson1974} for high energies, as it allows up to $10^8$ times faster simulations than MD \cite{mdbcacomp}, at the price of the loss of the time scale of the cascade. BCA codes can be divided into Monte Carlo based algorithm such as SRIM \cite{docsrim}, and deterministic codes such as MARLOWE \cite{robinson1992}.  SRIM was essentially designed to simulate amorphous media, due to the probabilistic estimation of the scattering and time integral for binary collisions. On the contrary, MARLOWE generates the crystalline structure of the material in a small neighbourhood of the followed particle \cite{robinson1992}, and can hence account for replacement sequences, focusons and channeling effects. 

The purpose of this work is to estimate the ballistic effects of ion beam irradiation on crystalline materials, using the ion mixing formalism. More specifically, the first objective is to show that BCA code MARLOWE can provide with reliable estimations of ballistic effects in pure copper, consistently with molecular dynamics for moderate energies, provided a rational parameter setting has been performed upstream. The second objective is to compute ballistic effects in the silver copper alloy under 1 MeV krypton ions, as such initial energy is beyond the scope of MD simulations. The general  motivation is the eventual development of multi-scale methods to link ion irradiation to patterned microstructure of materials at micron scale \cite{champ,champ2,demangethese2015,2016arXiv161001392D}. 

This study is structured as follows. First, the ion mixing formalism is briefly presented, and our rational parametrization of MARLOWE is detailed. Secondly, based on these settings, a first set of simulations is proposed, where the results from  our MARLOWE and MD simulations are compared, for a copper single crystal with moderate initial energies. This section allows to validate MARLOWE simulations, and provides with solid basis for the following simulations. Then, the case of AgCu immiscible alloy subjected to krypton ions (1 MeV) is studied using MARLOWE, as previous numerical \cite{Enrique2003} and experimental \cite{Wei1997} works suggested that ballistic displacements could lead to patterning effects in that case. Results are compared to experimental \cite{enrique2001} and numerical \cite{Enrique2003} studies, and then discussed from the angle of multi-scale studies on patterned microstructure induced by ion beam irradiation.

\section{Model \& Simulation method}
\label{sec:model}

\subsection{The ion mixing formalism}

Displacement cascades act on a very different time scale ($\sim 10^{-9}$ s) from microstructure diffusion dynamics ($\sim 10^{-6}$ s) \cite{demangethese2015}.   For a sufficient irradiation flux $\Phi$, numerous cascades hence occur during a fundamental time increment of the microstructure. This suggests to define an average cascade over this sample of single cascades. Only the effects of this virtual cascade impact the microstructure \cite{martin1984}. Albeit strongly heterogeneous when considered singly, due to subcascades spontaneous formation  above a certain threshold for the initial energy \cite{Walsh2013}, the average cascade can be considered homogeneous. This assumption requires that subcascades perform a covering of the average cascade's volume. The necessary condition for subcascades covering is that the estimate of the covering volume fraction $f_r$, exceeds the percolation threshold $P\sim 0.2$ in dimension 3 \cite{Misaelides1994}:

\begin{equation}
\label{percol}
f_r= n_{\text{subcasc}}\times \frac{l_{\text{subcasc}}}{L_{\text{casc}}}\geq P.
\end{equation}
Here, $n_{\text{subcasc}}$ is the average number of subcascades per cascade, and $L_{\text{casc}}$ and $l_{\text{subcasc}}$ are the average characteristic sizes of cascades and subcascades, respectively. For a spherical average cascade, $L_{\text{casc}}$ is the diameter of the circular section. By extension, for an ellipsoidal average cascade \cite{Misaelides1994},  $L_{\text{casc}}$ is defined correspondingly, from a virtual sphere of same volume as the cascade. The same definition goes for subcascades with $l_{\text{subcasc}}$. 

From this point, ballistic effects are only defined by an atomic relocation frequency $\Gamma$, and a probability distribution $p_R$ for an impinged atom to be relocated at a distance $r$ from its initial position. This probability $p_R$ is usually assumed to follow an exponential decay \cite{bellontheo}, associated to short range displacements within subcascades:

\begin{equation}
\label{loidep1}
p_R(r)=\frac{1}{4\pi R^3} \exp\left(\frac{-r}{R}\right).
\end{equation}
Here, $R$ is the mean relocation distance, associated to finite range disordering. As for $\Gamma$, it is the product of the irradiation flux $\Phi$, and a relocation macroscopic cross section $\sigma^d$ \cite{demangethese2015}.  $\sigma^d$ is defined as the characteristic surface $\pi (L_{\text{casc}}/2)^2$ of the cascade, weighed by an atomic displacement efficiency.  This relocation efficiency is the fraction of relocated atoms $N_d$, among all atoms in the volume of the cascade. $\Gamma$ thus eventually reads:    

\begin{equation}
\label{gamma}
\Gamma=\sigma^d \Phi, \quad \sigma^d=\frac{V_{\text{cell}}}{N_{\text{cell}}}\times \frac{N_d}{ L_{\text{casc}}},
\end{equation}
where $V_{\text{cell}}$ is the crystallographic lattice cell volume, and $N_{\text{cell}}$ is the number of atoms per lattice cell. 

This study focuses on the number of displaced atoms $N_d$, as well as the characteristic size of cascades $L_{\text{casc}}$ as they are both required to estimate the relocation frequency $\Gamma$, and the relocation  distribution of displaced atoms $p_R$. Particular attention is also paid to the percolation criterion \refeq{percol}, as required by the ion mixing theory.

\subsection{Parameter settings}

The first parameter governing collisions is the limit impact parameter $p_c$ for collision candidate atoms. In a pure binary collision approximation, this coefficient should ensure that collisions are isolated. Face centered cubic lattices require $p_c<0.35$ ($\times a$), corresponding to half distance between nearest neighbours, $a$ being the lattice parameter. Yet, this choice leads to overextended linear collision sequences \cite{robinson1974}, as $p_c$ is also involved in collision proposal wherein no interaction occur but electronic stopping. For a fcc lattice, the second closest neighbours at least must be taken into account, leading to $p_c=0.42$ ($\times a$). Note higher values for $p_c$ step up the occurrence of multiple collision, eventually  leading to an overestimation of the amount of relocated atoms by MARLOWE. Nevertheless, in case a replacement sequence is initiated, $p_c$ is then denoted $p_c^r$, and the neighbours among the $<011>$, $<001>$ and $<111>$ directions must be accounted for, so that  $p_c^r=0.62$. This choice allows the electronic stopping generated by neighbouring parallel atomic rows to help dissipate the linear collision sequence energy. 

\begin{figure}[ht]
\centering
\includegraphics[width=8cm]{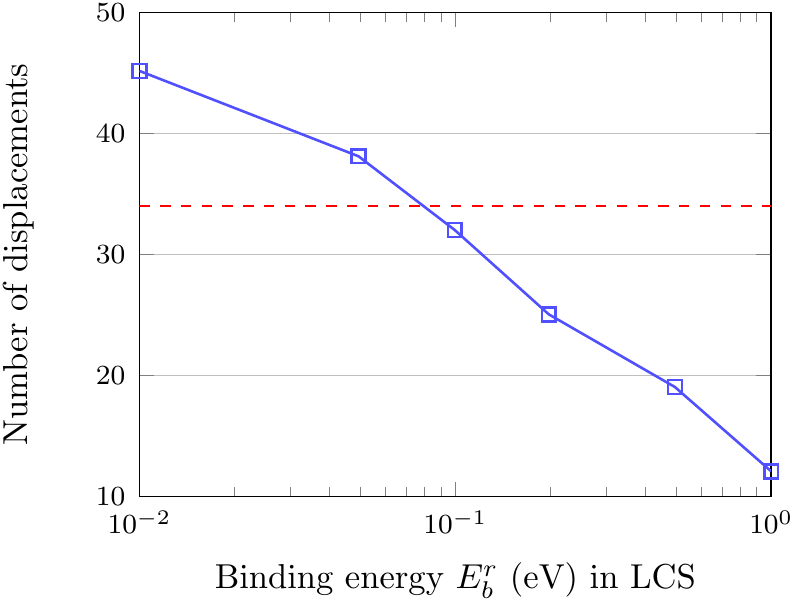}
\caption{Influence of the binding energy $E_b^r$ used in MARLOWE for linear collision sequences (LCS), on the number of displacements $N_d$ in the $<011>$ crystallographic direction (blue squares).  Sequence initiated with an initial energy of 25 eV and a cut-off energy $E_c\sim 0.05$ eV. Comparison with $N_d$ estimated by MD simulations in the same simulation conditions,  taken from \cite{coment} (red dashes).}
\label{fig:LCR} 
\end{figure}

The second parameter is the binding energy $E_b$ above which the target atom is set in motion, and a vacancy is created, albeit potentially unstable. The most physical choice is the enthalpy of vacancy formation of the material. This enthalpy is available in the literature for pure materials \cite{ebcuag}. In the case of alloys, the alloy formation energy must be added. These have been estimated via Modified Embedded Atom Method (MEAM) potentials \cite{ebcuco,eballiage}, or experimentally \cite{smithells}. However, this energy is overestimated in the case of linear collisions sequences (LCR) such as replacement sequences and focusons, as it is intuitively less expensive to switch atoms, than to eject an atom from its site thus leaving it vacant. Robinson suggested that this replacement binding energy should be about a tenth of the binding energy: $E_b^r\sim 0.1\times E_b$ \cite{robinson1974}. This assertion was checked in the $<011>$ crystallographic direction, as it is the most favourable direction for replacement sequences in fcc crystals. For that purpose, the number of replacements in this atomic row of copper was plotted in \reffig{fig:LCR}, for a 25 eV initial energy, as this small energy allows to get rid of electronic stopping. $E_b^r\sim 0.1 $  eV corresponding to $0.1\times E_b$ in the case of copper indeed led to the same number of displacements (34) as previous reliable MD simulations \cite{coment}.        

The cut-off energy $E_c$ is the threshold energy beneath which a moving atom stops. The wide range of values adopted in the literature \cite{cu1,mdbcacu,pred} suggests it is purely numerical. It corresponds to either the limit value beneath which the binary collision approximation is not valid, so that it should be set to 1-2 eV, or a threshold from which MD is used in the case of hybrid BCA-MD simulations \cite{hybrid}. In this study, $E_c$ was chosen as small as possible so that results were not degraded. Indeed, it was noted that when $E_c$ was set below 1.5 eV, the relocation distribution displayed an abnormal tail, and the ion penetration depth could diverge.

Simulations were performed on pure copper single crystal in a first time, then on a homogeneous silver-copper alloy single crystal. All useful physical and numerical parameters are displayed in table \ref{tab3}. 

\begin{table}[ht]
\centering
\begin{tabular}{ l l l }
   \hline
  & Cu (fcc) & Ag$_{50}$Cu$_{50}$ (fcc)\\
   \hline	
   \hline
$a$ (\AA{}) & 3.6015 & 3.844 \\
$N_{\text{cell}}$  & 4 & 4 \\
$V_{\text{cell}}$ (\AA{}$^3$) & 47.24 & 56.80 \\
   \hline	
   \hline
$E_b$ (eV) & 1.10 & 1.16--1.26 \\
$E_b^r$  (eV)  & 0.1 & 0.12--0.13 \\
$E_c$  (eV)  & 1.5 & 1.5  \\
$p_c$ ($\times a$) & 0.5 (0.42\footnote{$E_{\text{PKA}}\geq 10$ keV}) & 0.5 (0.42\footnotemark[1])\\
$p_c^r$ ($\times a$) & 0.62 & 0.62 \\
   \hline

\end{tabular}

\caption{MARLOWE parameters for collision description used in this study for copper single crystal, and silver-copper homogeneous single crystal. $a$: lattice cell parameter, $N_{\text{cell}}$: number of atoms per unit lattice cell, $V_{\text{cell}}$: lattice cell volume, $E_b$: binding energy, $E_b^r$: binding energy for replacement sequences, $E_c$: cut-off energy, $p_c$: limit impact parameter, $p_c^r$: limit impact parameter for replacement sequences.}
\label{tab3}
\end{table}

As regards the inelastic potential $S_e$, MARLOWE's native potential derives from the Lindhard, Scharff and Schi\o tt (LSS) theory  \cite{lss1}. Nevertheless, this potential is only valid for energies below a few keV by atomic mass unit, as it does not account for Bethe effects that arise for primary knocked atoms (PKA) energy exceeding 100 keV \cite{bethe}. It was hence necessary to modify $S_e$ in this study, as 1 MeV krypton ions were used. SRIM potentials were transferred to MARLOWE, as they were determined precisely to ensure a reliable computation of the ion implantation peak \cite{docsrim}. Finally, the standard Moli\`erere potential for the nuclear stopping power was used \cite{melker2009potentials}. 

As for the details on MD parameters setting and computations, they have already been presented and discussed elsewhere \cite{Antoshchenkova2015168}.

\section{Results and Discussions}
\label{sec:results}

\subsection{First MARLOWE simulations on pure copper \& comparison with MD}
\label{subsec:valid}

Here we present and compare the results from our BCA and MD calculations. This provides with a solid basis for the high energy cascade simulations of the following section, where MD simulations cannot be performed. In this section, all simulations were realised on an infinite single crystal, subjected to volume irradiation. 

\begin{figure}[ht]
\centering
\includegraphics[width=8cm]{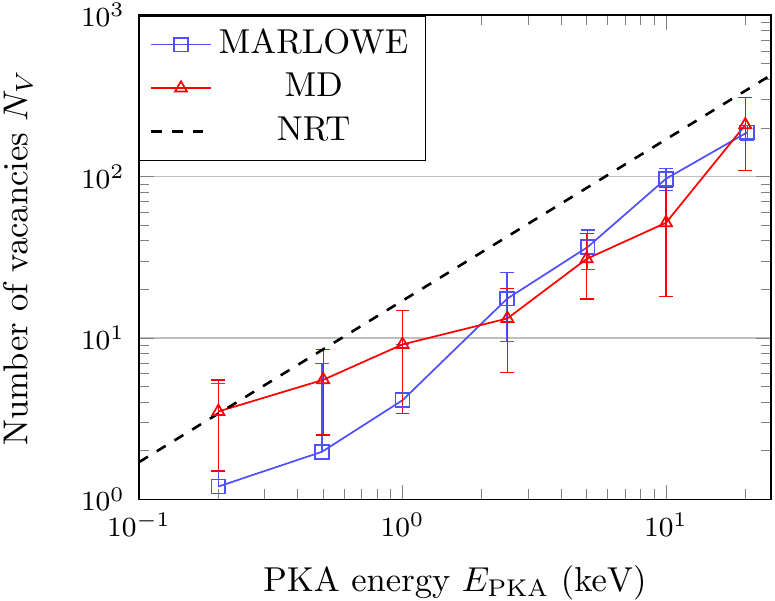}
\caption{Number of vacancies $N_V$ in copper single crystal, for a primary knocked on copper atom (PKA) of energy $E_{\text{PKA}}$ ranging from 0.5 keV to 20 keV. Comparison between our MARLOWE simulations averaged over 1000 trials (square), our MD simulations averaged over 10 trials (triangle), and NRT law (dashes).}
\label{fig:Nv} 
\end{figure}

Preliminary to ballistic effects, the number of Frenkel pairs from BCA and MD simulations are prospected. \reffig{fig:Nv} displays the numbers of point defects $N_V$ obtained for different PKA energies by MARLOWE (square) and MD (triangle) methods. A good agreement was found with the exception of low energies. This discrepancy of BCA simulations for low energies is due to the predominance of relaxation mechanism over binary collisions \cite{Koponen1997}. It is important to note that in both MARLOWE and MD simulations, the dependence of $N_V$ toward incident energy  is similar to the Norton-Robinson-Torrens (NRT) analytical law \cite{was2007} (dashed line in \reffig{fig:Nv}). Besides, the overestimation of $N_V$ by the NRT law is recovered \cite{nrt}. 

\begin{figure}[ht]
\centering
\subfigure[~~~MARLOWE]{\includegraphics[width=4.2cm]{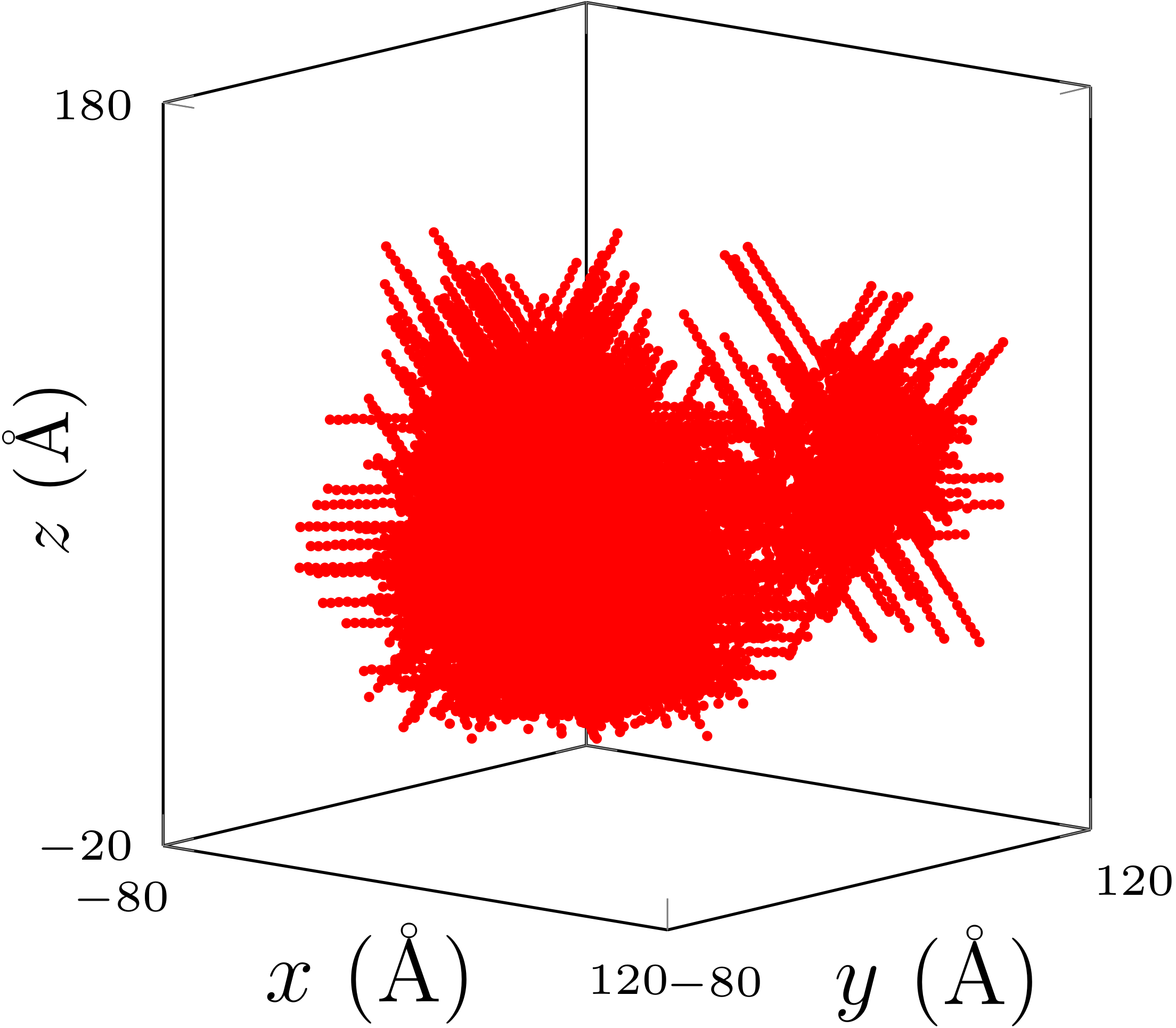}\label{fig:cascbca}}
\subfigure[~~~MD]{\includegraphics[width=4.2cm]{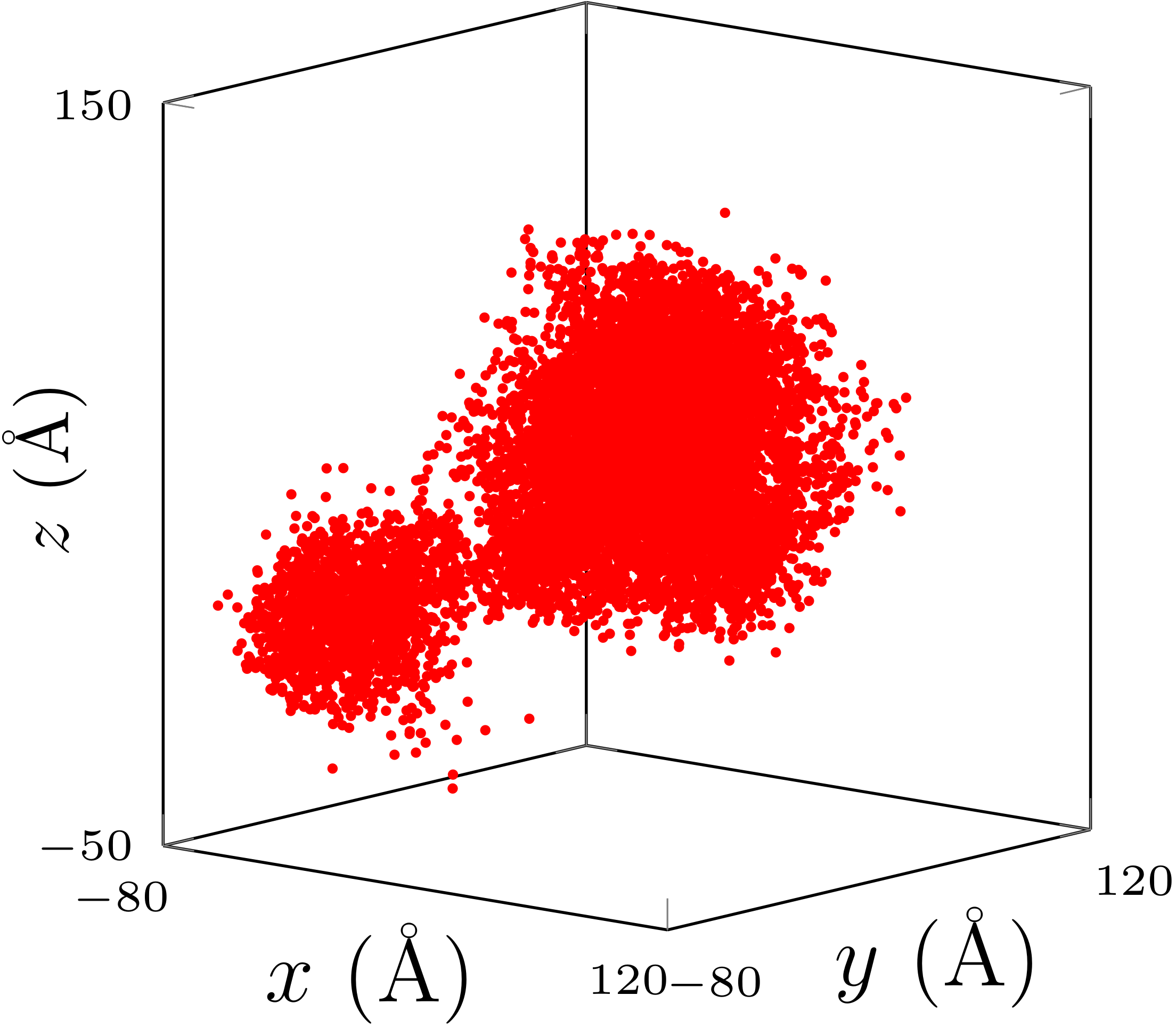}\label{fig:cascmd}}
\caption{Displacement cascades in single crystal copper for a primary knocked on copper atom (PKA) of energy $E_{\text{PKA}}=20$ keV. Position of relocated atoms before recombination process are presented. Left: from 1 random trial of our MARLOWE simulations. Right: from 1 random trial of our MD simulations. The cascade can be divided into 2 subcascades in both cases. }
\label{fig:casccu} 
\end{figure}

Then, table \ref{tab1} displays the characteristic size of simulated cascades $L_{\text{casc}}$, required to estimate $\Gamma$ using \refeq{gamma}. $L_{\text{casc}}$ increases with incident energy, showing good consistency with our MD estimation, as all values are within 15 \% disparity. The number of subcascades $n_{\text{subcasc}}$ is also displayed in table \ref{tab1}, as it is required in \refeq{percol}. In particular, 2 subcascades are formed for $E_{\text{PKA}}=20$ keV in  both MARLOWE and MD simulations, as displayed in \reffig{fig:casccu}. It is interesting to note that this result confirms the ability of BCA simulations to account for subcascades formation under ion irradiation \cite{FUKUMURA1991}. Besides, the shape of the global cascade obtained with MARLOWE in \reffig{fig:cascbca} is very similar to the cascade simulated in the same conditions with molecular dynamics in \reffig{fig:cascmd}. More generally, no subcascades were observed beneath 20 keV in MARLOWE and MD simulations. This is consistent with both theoretical works on the fractal description of cascades \cite{kelly1989,Luneville201155}, and numerical studies on the PKA energy threshold for the emergence of subcascades in metals \cite{Metelkin1997,Walsh2013}. This was indeed estimated to be 10-20 keV in copper \cite{Antoshchenkova2015168}.   

\begin{table}[ht]
\centering
\begin{tabular}{ l l l l l l l}
   \hline
$E_{\text{PKA}}$ (keV)  & \multicolumn{2}{c}{$n_{\text{subcasc}}$} & \multicolumn{2}{c}{$L_{\text{casc}}$ (\AA{})} \\
 & MD & MARLOWE & MD & MARLOWE \\
   \hline	
   \hline
1     & $\varnothing$ & $\varnothing$ &  33 $\pm 18$  & 28 $\pm 10$\footnote{Indicative standard deviation, depending on rare events truncation.} \\
5     & $\varnothing$ & $\varnothing$ &  78 $\pm 8$   & 70 $\pm 10$\\
10    & $\varnothing$ & $\varnothing$ &   97 $\pm 28$  & 90 $\pm 10$\\
20    & 2 & 2 &  142 $\pm 23$ & 139 $\pm 10$\\
   \hline

\end{tabular}

\caption{Number of subcascades $n_{\text{subcasc}}$, and mean length of cascades $L_{\text{casc}}$, as a function of PKA energy $E_{\text{PKA}}$, for copper single crystal, averaged over 1000 trials for MARLOWE, and 10 trials for MD.}
\label{tab1}
\end{table}

The number of subcascades $n_{\text{subcasc}}=2$, and the characteristic size of subcascades $l_{\text{subcasc}}\sim 50$ \AA{}, and of cascades $L_{\text{casc}}\sim 139$ \AA{}, allows the computation of the covering volume fraction $f_r$  for $E_{\text{PKA}}=20$ keV. \refeq{percol} provided with  $f_r\simeq 0.5$. And so, $f_r$ exceeds the percolation threshold $P=0.2$, and the percolation criterion in \refeq{percol} validates the use of the ion mixing framework. For lower values of $E_{\text{PKA}}$, no clear subcascades were observed. The cascade could be considered as one subcascade occupying the overall volume of the cascade. The percolation threshold was then automatically reached, as $n_{\text{subcasc}}$ =1, $l_{\text{subcasc}}/L_{\text{casc}}=1$, so that $f_r=1$.

\begin{figure}[ht]
\centering
\includegraphics[width=8cm]{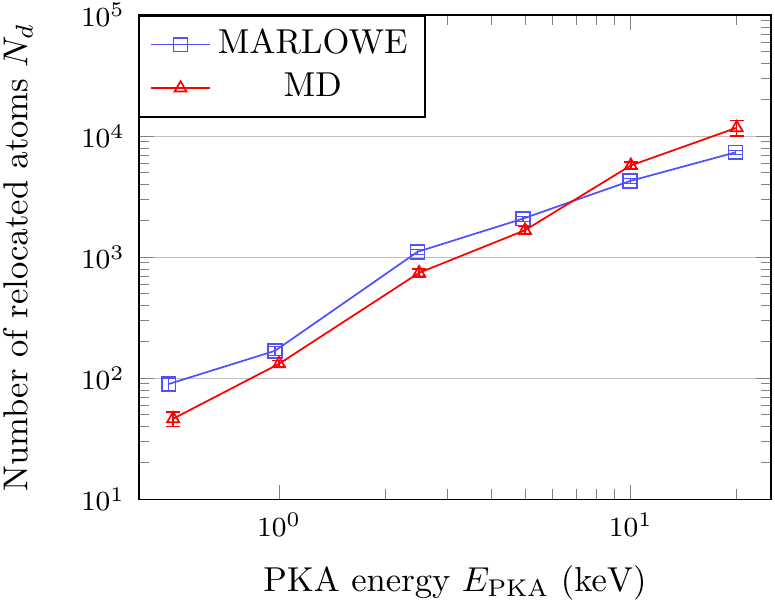}
\caption{Number of relocated atoms $N_d$ in copper single crystal, for a primary knocked on copper atom (PKA) of energy $E_{\text{PKA}}$ ranging from 0.5 keV to 20 keV. Comparison between our MARLOWE simulations averaged over 1000 trials (square), and our MD simulations averaged over 10 trials (triangle).}
\label{fig:Nd} 
\end{figure}

The core result of the study is the computation of the relocation frequency $\Gamma$ in \refeq{gamma}, and the relocation probability $p_R$ in \refeq{loidep1}. As for $\Gamma$ first, its estimation requires the total amount of relocated atoms per cascade $N_d$, along with $L_{\text{casc}}$. As displayed in \reffig{fig:Nd}, both simulation method showed that $N_d$ increases with $E_{\text{PKA}}$. Besides, MARLOWE and MD simulations match, except for small energies ($<1$ keV), once again due to the loss of accuracy of BCA codes for small energies. A disparity between MARLOWE and MD is also observed for 20 keV. This is possibly the thermal spike contribution to displacements, which is characterized by short range movements of a large number of atoms within a roughly spherical volume, and cannot be modeled within the BCA framework. Given an irradiation flux $\Phi$, $\Gamma$ could then be computed through the relocation cross section $\sigma^d$. For $E_{\text{PKA}}=5$ keV, 10 keV and 20 keV, expression \refeq{gamma} provided with $\sigma^d=0.35\times 10^{-13}$ cm$^2$, $\sigma^d=0.55\times 10^{-13}$ cm$^2$ and $\sigma^d=0.62\times 10^{-13}$ cm$^2$, respectively. For an experimentally reachable flux $\Phi=10^{13}$ cm$^{-2}$s$^{-1}$, MARLOWE provides with  $\Gamma=\sigma^d \Phi=0.35$ s$^{-1}$, $\Gamma=0.55$ s$^{-1}$ and $\Gamma=0.62$ s$^{-1}$, respectively. For the same flux and energies, MD provides with $\Gamma=0.25$ s$^{-1}$, $\Gamma=0.69$ s$^{-1}$ and $\Gamma=0.96$ s$^{-1}$, respectively. MARLOWE and MD simulations hence provide with compatible estimations of $\Gamma$.

\begin{figure}[ht]
\centering
\includegraphics[width=8cm]{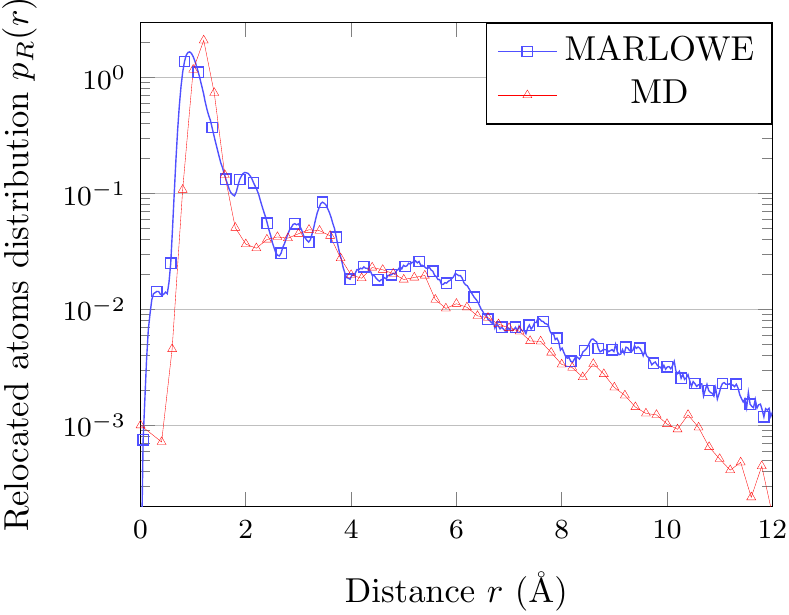}
\caption{Relocation distribution $p_R(r)$ in copper single crystal for a primary knocked on copper atom  (PKA) of energy $E_{\text{PKA}}=20$ keV. Comparison between our MARLOWE simulations averaged over 1000 trials (square), and our MD simulations averaged over 10 trials (triangle). In both cases, $p_R(r)$ displays an exponential decay.}
\label{fig:depcu} 
\end{figure}

As for the relocated atoms distribution $p_R$, it is displayed in \reffig{fig:depcu} for MARLOWE and MD calculations, using $E_{\text{PKA}}=20$ keV. The choice of maximal interval of the depth considered for the exponential regression was limited by our MD simulations.  For instance, for $E_{\text{PKA}}=20$ keV, we used  relocation distances comprised between 1.8 \AA{} and 13 \AA{}, leading to a 0.98 regression reliability. Both MARLOWE (square) and MD (triangle) distributions can be considered as an exponential decay, as displayed in \reffig{fig:depcu}. The exponential decay differs from the standard Levy flight model \cite{simeone2010b}, as the latter also accounts for long displacements between subcascades. The same conclusion applies to all initial energies. It is interesting to note the succession of periodic peaks for both distributions, at $r\simeq 1.2$ \AA{}, $r\simeq 2.2$ \AA{} and $r\simeq 3.6$ \AA{} in $p_R$. This traduces the preference for relocated atoms to move from a site to one of the neighbouring sites \cite{Enrique2003}.

\begin{figure}[ht]
\centering
\includegraphics[width=8cm]{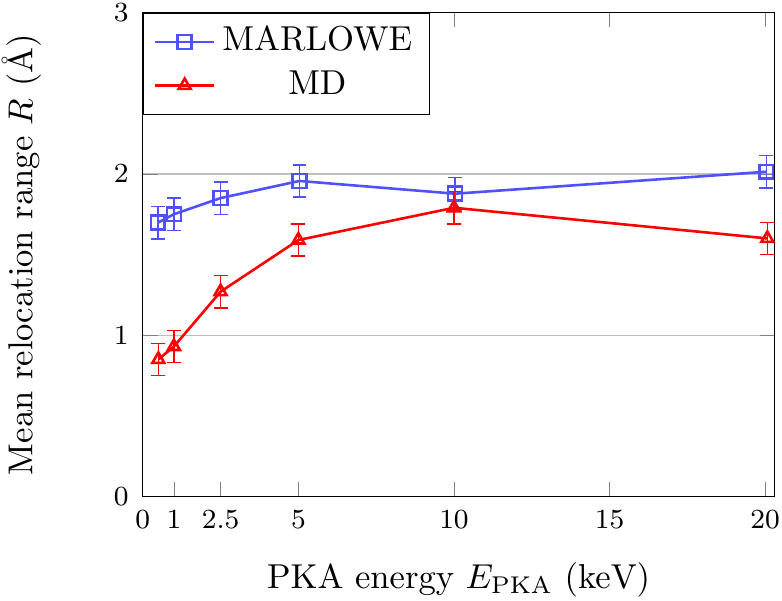}
\caption{Mean relocation range $R$ in the exponential decay probability in copper single crystal, for a primary knocked on copper atom (PKA) of energy $E_{\text{PKA}}$ ranging from 0.5 keV to 20 keV. Comparison between our MARLOWE simulations averaged over 1000 trials (square), and our MD simulations averaged over 10 trials (triangle).}
\label{fig:R} 
\end{figure}

The exponential decay law for relocation distributions is characterized by the mean relocation range $R$, which corresponds to the the average mixing length of ballistic effects. It is graphically defined as the inverse of the slope in semi-logarithmic scale. It is displayed in \reffig{fig:R} from our MARLOWE (square) and MD (triangle) simulations, for various PKA energies. $R$ increases with $E_{\text{PKA}}$ for small energies ($E_{\text{PKA}}\leq 2.5$ keV), and eventually converges to an asymptotic value $R\simeq 1.9\pm 0.1$ \AA{} for MARLOWE, and $R\simeq 1.7\pm 0.1$ \AA{} for MD (average over 5 keV, 10 keV and 20 keV). It can be argued that this asymptotic value of $R$ correctly describes the exponential decay for sufficient energies, \emph{i.e.} where the ion mixing is valid. Finally, on this range of energies, $R$ can be considered as constant, for a given ion-crystal target couple. The same conclusion goes for the relocation distribution $p_R$.

\subsection{Application to the AgCu system}
\label{subsec:valid}

The silver copper alloy subjected to higher energies was experimentally \cite{bellon2001} and numerically \cite{bellontheo} studied. Contrary to the previous section, the energies involved here are out of reach of our MD simulations, so that only MARLOWE simulations are available. Yet, the parametrization of MARLOWE, previously validated with MD, remains the same. In these simulations, a semi-infinite homogeneous AgCu single crystal was irradiated with krypton ions of energy $E_{\text{inc}}=1$ MeV. The external surface of the target was oriented perpendicularly to the (Oz) axis, and the crystal was oriented correspondingly. Besides, the initial velocity of incident ions was always along the (Oz) axis, but the initial impact position was chosen randomly. The presented results were obtained after an average over 1000 random trials. 

\begin{figure}[ht]
\centering
\includegraphics[width=8cm]{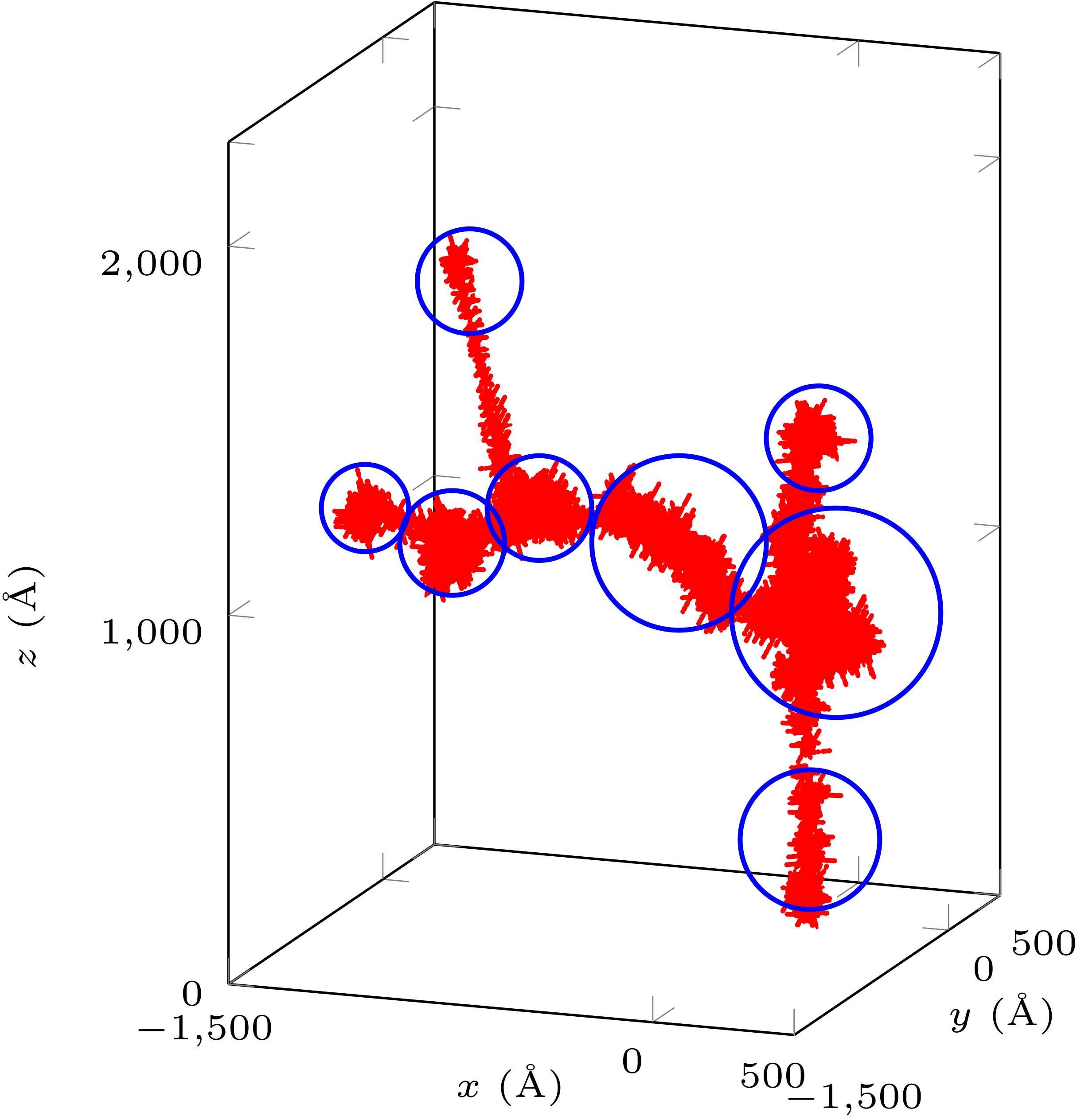}
\caption{Displacement cascade in  silver-copper single crystal for krypton  incident ions ($E_{\text{inc}}=1$ MeV), from 1 random trial of our Marlowe simulations.  Positions of relocated atoms before recombination process are presented. The cascade can be divided into 8 subcascades (circles).}
\label{fig:cascagcu} 
\end{figure}

The first step is the validation of the percolation criterion \refeq{percol}, through the morphology of displacement cascades (\reffig{fig:cascagcu}). The average characteristic size is $L_{\text{casc}}\simeq 210$ nm. Contrary to lower energies as in the previous section, the averaged displacement cascade is not spherical, and $L_{\text{casc}}$ cannot be associated to the length of cascade nor the ion penetration range. It is rather derived from the ellipsoidal estimation of the average cascade volume \cite{Misaelides1994}. Within these cascades, an average of $n_{\text{subcasc}}\sim 6$ subcascades was observed (8 subcascades in \reffig{fig:cascagcu}). This amount of subcascades is of consistent order with \cite{Metelkin1997}, as PKA energies ranged from 80 keV to 180 keV for copper and silver atoms in our simulations. Moreover, their mean size is at least $l_{\text{subcasc}}\geq 15$ nm, eventually leading to a covering fraction $f_r\geq 0.4$, sufficient to ensure subcascade covering.

Then, the total amount of relocations per cascade was found to be $N_d=309 000$, which is slightly inferior to \cite{Enrique2003}, as they found $N_d=421000$. This difference is partly due to non ballistic relocations, left aside by BCA simulations. It echoes the slight underestimation of $N_d$ by MARLOWE compared to MD, that was observed in the case of pure copper in the previous section. Together with $L_{\text{casc}}\simeq 210$ nm, this allowed the computation of the relocation frequency $\Gamma$ for any irradiation flux, through the relocation cross section $\sigma^d=2.01\times 10^{-13}$ cm$^2$. The irradiation of AgCu by 1 MeV krypton ion was already performed experimentally in \cite{bellon2001}, with $\Phi=1.21\times 10^{13}$ cm$^{-2}$s$^{-1}$, providing them with $\Gamma=2.8$ s$^{-1}$. We used the same irradiation flux. $\Gamma=\sigma^d \Phi$ was eventually found equal to 2.53 s$^{-1}$, therefore showing good consistency.

\begin{figure}[ht]
\centering
\includegraphics[width=8cm]{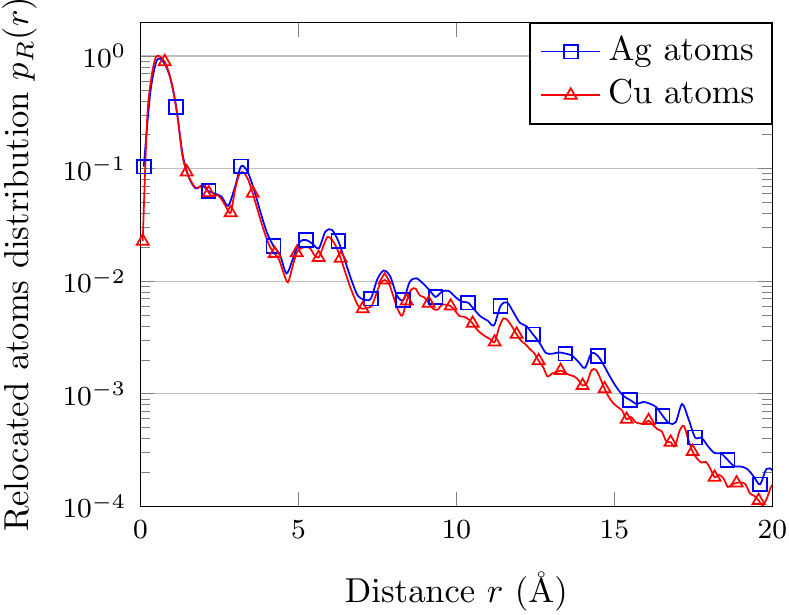}
\caption{Relocation distribution (normalized) $p_R(r)$ of silver (square) and copper (triangle) atoms in AgCu semi-infinite single crystal, for krypton incident ions ($E_{\text{inc}}=1$ MeV), from our MARLOWE simulations averaged over 1000 trials. The relocation distribution  $p_R(r)$ for copper and silver atoms can be fitted by a unique exponential decay ($\rho=0.98$ for $r\in[2,20]$ \AA{}), with $R=3.04\pm 0.01$ \AA{}.}
\label{fig:distribagcu} 
\end{figure}

Then, \reffig{fig:distribagcu} displays the relocation distribution of silver (square) and copper (triangle) atoms, for a flight distance ranging from 2 \AA{} to 20 \AA{}. Longer displacements are of negligible contribution to $p_R$ \cite{Enrique2003}, as for $r>20$ \AA{}, $p_R<10^{-4}$ corresponds to less than 30 relocations in the cascade. Once again, the exponential decay law is satisfied ($\rho=0.98$). Besides, both distributions are very close, so that it can arguably be described by a unique exponential decay, of corresponding mean relocation range $R=3.04$ \AA{}.

\begin{figure}[ht]
\centering
\includegraphics[width=8cm]{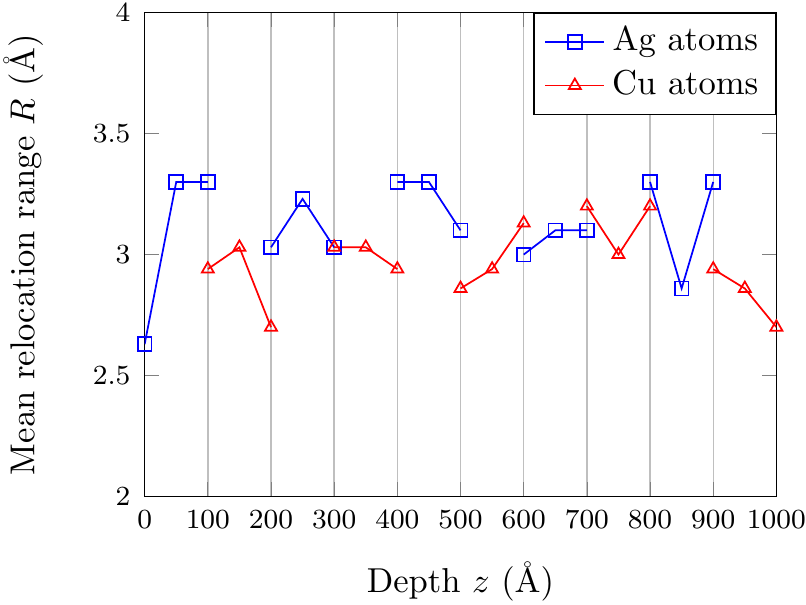}
\caption{Mean relocation range $R$ of the exponential decay probability as a function of depth for silver (square) and copper (triangle) atoms, in a semi-infinite multi-layer material composed of silver and copper single crystals layers (width$=10$ nm). $R$ is determined from the relocation distribution of displaced atoms initially in a given layer. $R$ varies very little with depth. Average on 1000 trials of our MARLOWE simulations, with krypton incident ions ($E_{\text{inc}}=1$ MeV) .}
\label{fig:multi} 
\end{figure}

As part of the study of the emergence of patterned microstructure in irradiated binary alloys, one might question the relevance of a homogeneous 
single crystal to simulate ballistic effects on the microstructure. In the case of AgCu for instance, the patterned microstructure will display silver and copper enriched regions, separated by semi coherent interfaces, due to the 13\% misfit between pure copper and pure silver lattice cell parameters \cite{shao2013}. Two difficulties hence arise: the change of the crystal bulk property from AgCu solid solution, and the creation of dislocations at the interfaces between enriched domains. The change of crystal bulk properties should theoretically change the binding energies used in table \ref{tab3}. Here, Cu, Ag and AgCu display similar binding energies \cite{Bandyopadhyay1990,Nordlund1998,McGervey1975}. As for the impact of dislocations at the interface between enriched regions, we suspected it would impact the relocation distribution $p_R$, due to unchanneling effects in particular. To challenge this hypothesis, irradiation effects in a characteristic labyrinthine pattern was modeled by a multilayer target, composed of 10 nm layers of silver and copper single crystals, alternatively disposed, with the same crystal orientation in the first place. The same irradiation simulation as before (1 MeV krypton incident ions) was used, and  $R$ was plotted as a function of the initial depth of relocated atoms in \reffig{fig:multi}.

Two interesting observations can be made. First, despite small fluctuations, $R$ remains very close to a mean value over the whole crystal depth: $R=3.05\pm 0.08$ \AA{}. This proximity comes from the fact that the dislocations at the interface mainly break replacement sequences, and channeling displacements. The disruption of replacement sequences have almost no impact on $p_R$. Indeed, replacement sequences involve extremely short range displacements (1-2 \AA{}), which are not taken into account in the relocation distribution fit. As for channeled atoms, they remain quite scarce ($<1$ \textperthousand{} of all displacements), and have no significant impact on $p_R$.  It is important to note that the crystallographic disorientation of layers has no impact on $\Gamma$ nor  $p_R$, provided the width of layers is not too small. When the width becomes too small, the multilayer material behaves like an amorphous material \cite{robinson1992}. Second, one might expect the mean value of $R$ of the multilayer material, to be comprised between that of pure copper ($R\sim 1.9$ \AA{}) and pure silver ($R\sim 2.5$ \AA{} from MARLOWE simulations not presented here). Actually, it is very close to the homogeneous AgCu alloy value ($R=3.04$ \AA{}). We believe this is a statistical effect. The average effects of randomly distributed subcascades on the multilayer material could hence be the same as the effect of a cascade on the disordered alloy, provided the layers of the multilayer material are thin enough. Nevertheless, the interruption of linear collision sequences results in a decrease of the number of displaced atoms ($\sim -20$ \%). The estimation of $\Gamma$ using an homogeneous crystal model might hence be  slightly overestimated compared to the real patterned material, as modeled in \reffig{fig:multi}. Apart from this bias, ballistic effects estimation in homogeneous polycrystalline AgCu alloy are close enough to the multilayer material to be transferred to heterogeneous patterned microstructure in AgCu.

\section{Conclusion}

In this study, the ballistic effects of ion beam irradiation on metals were investigated within the ion mixing framework, using the Binary Collision Approximation code MARLOWE. Two sets of collision cascades simulations were presented. 

In order to validate MARLOWE's parametrization, the first set of simulations was performed using both MARLOWE and Molecular Dynamics, on a pure copper single crystal, for PKA energies ranging from 0.5 keV to 20 keV. All cascade features needed in the ion mixing formalism were investigated. The percolation threshold was checked to be reached. Then, the exponential decay for the displacement probability $p_R$ was confirmed, and the mean relocation range $R$ was computed. Finally, the relocation frequency $\Gamma$ was also estimated. Comparison between BCA and MD results showed very good consistency for both morphology and average features of the cascades. 

Since the ion mixing is widely used to study the microstructuration of irradiated alloys, the homogeneous silver copper alloy was chosen for the study, in the second set of simulations. The validated parametrization of MARLOWE was used with 1 MeV krypton incident ions, as numerical and experimental results were available in the literature to validate our simulations. Once again, results showed excellent consistency.

The two main results of this study are the following:

\begin{itemize}

\item MARLOWE was rationally parametrized, proved to be predictive, and provided with results that were consistent with MD simulations.

\item Ballistic effects in AgCu alloy under 1 MeV krypton ions were computed with MARLOWE, within the ion mixing framework, so that the first step toward microstructuration of irradiated metallic alloys was achieved.

\end{itemize}

%

\end{document}